\documentclass[a4paper,10pt,twocolumn]{article}
\usepackage[top=2.5cm, bottom=2.5cm, left=2.1cm, right=2.1cm]{geometry}
\usepackage{color,soul}
\usepackage[utf8]{inputenc}
\usepackage{graphicx}
\usepackage{cite}
\usepackage{lineno}
\title{Ultra-broadband nanophotonic beamsplitter using an anisotropic sub-wavelength metamaterial}
\author{Robert Halir$^{1,2,*}$, Pavel Cheben$^{3}$, José Manuel Luque-González$^{1}$,  \\ Jose Darío Sarmiento-Merenguel$^{1}$, Jens H. Schmid$^{3}$, Gonzalo Wangüemert-Pérez$^{1}$, \\Dan-Xia Xu$^{3}$, Shurui Wang$^{3}$,  Alejandro Ortega-Moñux$^{1,2}$, and Íñigo Molina-Fernández$^{1,2}$ \\ \\ $^1$Universidad de Málaga, Dept. de Ingeniería de Comunicaciones, ETSI \\ Telecomunicación, Campus de Teatinos s/n, 29071 Málaga, España\\ $^2$Bionand Center for Nanomedicine and Biotechnology,\\ Parque Tecnológico de Andalucía, 29590 Málaga, España\\ $^3$National Research Council of Canada, Ottawa, Ontario, K1A 0R6, Canada \\ $^*$Corresponding author: robert.halir@ic.uma.es}
\date{}
\begin{document}

\maketitle

%\begin{abstract}
\textbf{Nanophotonic beamsplitters are fundamental building blocks in integrated optics, with applications ranging from high speed telecom receivers to biological sensors and quantum splitters. While high-performance multiport  beamsplitters have been demonstrated in several material platforms  using multimode interference couplers,  their operation bandwidth remains fundamentally limited.  Wavelength independent integrated beamsplitters would enable breakthrough broadband integrated photonic systems  for communications, sensing and quantum physics. Here, we leverage for the first time the inherent anisotropy and dispersion of a sub-wavelength structured photonic metamaterial to demonstrate  ultra-broadband integrated beamsplitting. Our device,  which is three times more compact than its conventional counterpart, can achieve virtually perfect operation over an unprecedented $\mathbf{500\,nm}$ design bandwidth exceeding all optical communication bands combined, and making it the most broadband silicon photonics component reported to date.  Our demonstration paves the way toward  nanophotonic waveguide components with ultra-broadband operation for next generation integrated photonic systems. 
}
%\end{abstract}

\begin{figure}[tbp]
\centering
{\includegraphics[width=\linewidth]{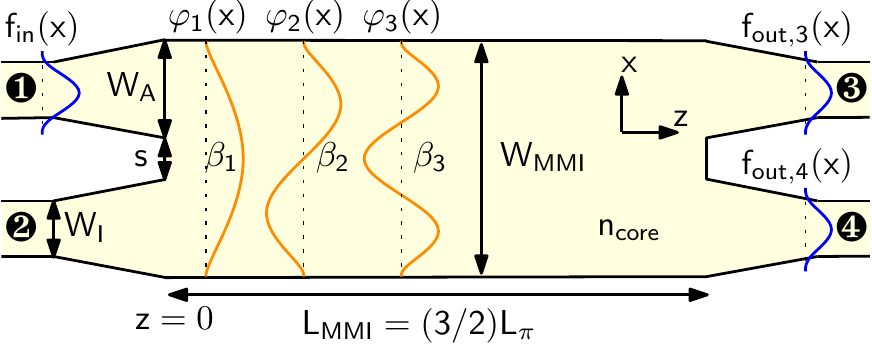}}
\caption{Schematic 2D model of a $2\times 2$ MMI. When the mode of the input waveguide, $f_\mathrm{in}(x)$, is launched into the multimode region it excites several higher order modes, $\varphi_i(x)$, which interfere forming Talbot-type self-images of the input field. Coupling of these self-images to the output waveguide yields the output fields $f_{\mathrm{out},3}(x)$ and $f_{\mathrm{out},4}(x)$, each carrying half of the input power with a relative phase shift of $90^\circ$. 
%$f_{\mathrm{out},3}(x)=f_\mathrm{in}(x)/\sqrt{2}$ and $f_{\mathrm{out},4}(x)= f_\mathrm{in}(x)\exp(\mathrm{j}\pi/2)/\sqrt{2}$.
}
\label{fig:mmi_conv}
\end{figure}

Integrated photonic systems are poised to produce  key advances in areas such as optical communications \cite{dong2014monolithic,sano2012102}, sensing \cite{estevez2012integrated, rodrigo2015mid}, spectroscopy \cite{nedeljkovic2016mid, le2007wavelength}, metrology \cite{pasquazi2011sub}, frequency comb generation \cite{razzari2010cmos} and quantum physics \cite{crespi2013anderson, he2015ultracompact}. Since beamsplitters are fundamental photonic building blocks, substantially extending their operational bandwidth would pave the way towards broadband systems capable, for instance, of covering several optical communication bands, or realizing sensors and frequency combs over large spectral regions as required in optical coherence tomography. Integrated beamsplitters are preferably implemented  with multimode interference couplers (MMIs), which exploit the self-imaging effect discovered by Henry Talbot in the 1830s \cite{talbot1836}. Talbot imaging allows to form multiport couplers with well defined amplitude and phase ratios that offer improved performance compared to y-branches and directional couplers \cite{ulrich1975, soldano1995optical}. MMI couplers  are used in a wide variety of waveguide devices  \cite{dong2014monolithic,he2015ultracompact,runge2012monolithic,Halir:11,Zhang:13, Kleijn,Fandino:15}, and have been implemented in a broad range of material platforms \cite{TajaldiniNonLinear,zhu2013compact,guan2014extremely,ZhengGraphene}. 
However, until very recently, the same classical structure, similar to the one shown in Fig. \ref{fig:mmi_conv} for a $2\times 2$ MMI configuration, has been used.  Consisting of an essentially rectangular central multimode region with tapered input and output waveguides, it provides a rugged design, but also imposes fundamental restrictions in terms of  size and bandwidth \cite{besse1994optical,soldano1995optical}. 
On the other hand, the recent advent of sub-wavelength grating (SWG) metamaterial  waveguides, which suppress diffraction effects,  has enabled the concept of refractive index engineering \cite{cheben2010refractive, halir2015waveguide}. This concept has ushered in  breakthrough devices for fiber-to-chip coupling \cite{Cheben:15,Benedikovic:15},  wavelength multiplexing \cite{piggott2015inverse}, and polarization splitting \cite{shen2015integrated}, among other innovations. Applied to MMIs, sub-wavelength index engineering led to the first demonstration of a slotted MMI \cite{ortega2013ultra}, which reduced the size of the device by a factor of two compared to a conventional MMI, without affecting its performance. 

\begin{figure*}[tbp]
\centering
{\includegraphics[width=0.95\linewidth]{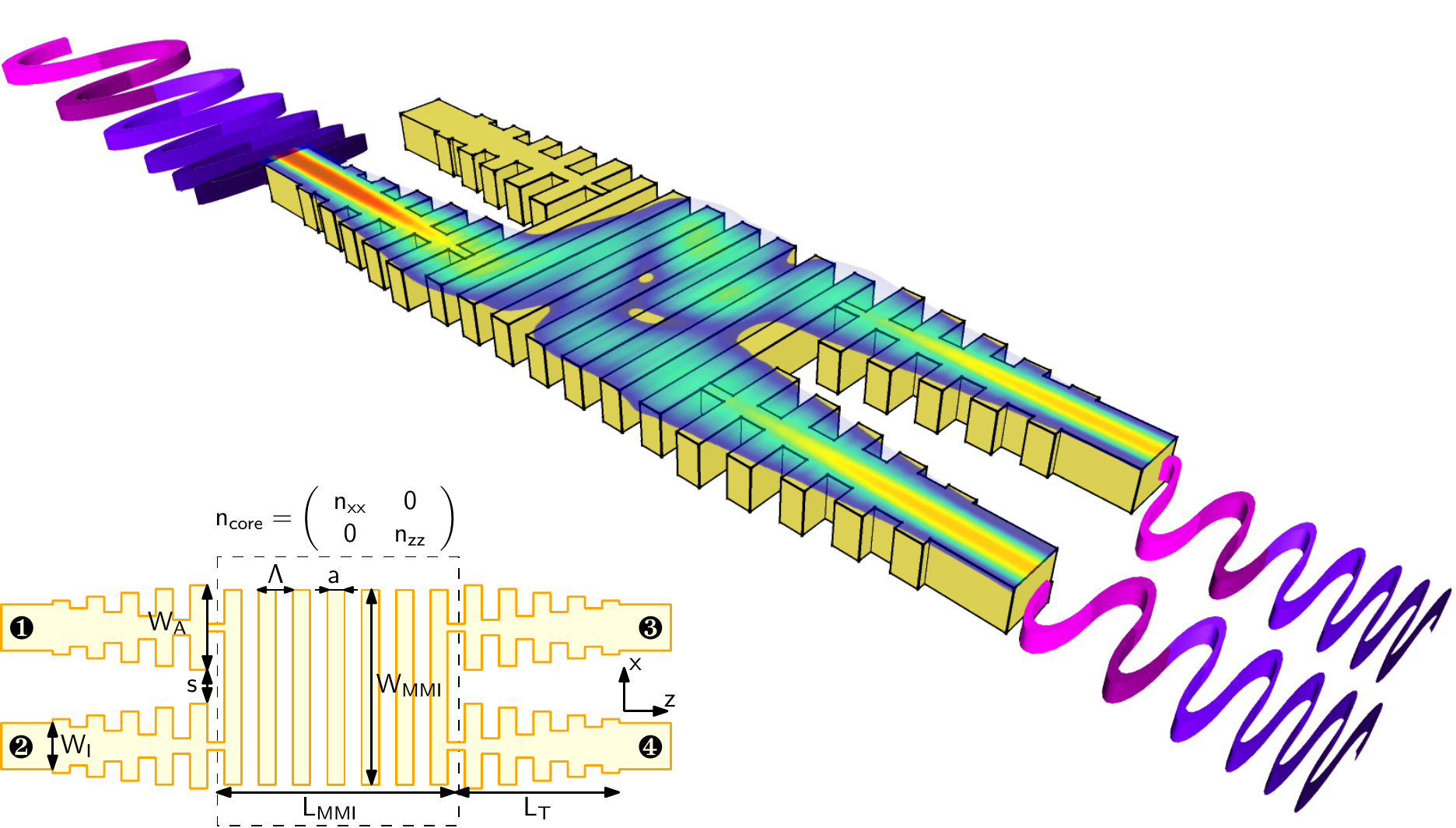}}
\caption{Ultra-broadband multimode interference coupler. The central multimode region is segmented at a sub-wavelength scale to engineer the waveguide anisotropy and dispersion, achieving a beat length that is virtually independent of wavelength. The sub-wavelength grating tapered input and output waveguides ensure efficient coupling with the silicon wire interconnecting waveguides. In the illustration TE polarized light (electric field along the $x$-axis) with varying wavelength is injected into one of the input ports and is split among the two output waveguides with equal amplitude and a $90^\circ$ phase shift.}
\label{fig:mmi3d}
\end{figure*}

Here we examine, for the first time, self-imaging in an anisotropic sub-wavelength structure and exploit this new concept, in combination with dispersion engineering \cite{maese2013wavelength}, to experimentally demonstrate a compact and ultra-broadband MMI (see Fig. \ref{fig:mmi3d}). 
Our fabricated device yields a threefold size reduction compared to a conventional device, while at the same time achieving perfect performance (excess losses and imbalance $<1\,\mathrm{dB}$,  phase error $<5^\circ$) over an unprecedented bandwidth of more than $300\,\mathrm{nm}$ around a central wavelength of $1.55\,\mu\mathrm{m}$, limited by our measurement setup. 
Full 3D simulations predict the bandwidth exceeding  $500\,\mathrm{nm}$; in simulation an optimized conventional MMI design covers less than $200\,\mathrm{nm}$ with comparable performance.
To the best of our knowledge this not only constitutes the MMI with the broadest bandwidth ever demonstrated, but also one of the most broadband, fully passive integrated optical devices in general.    

%The manuscript is organized as follows. In section 2 we review the operation of conventional MMIs and present a semi-analytical 2D model for the sub-wavelength patterned MMI, based on multimode propagation in an anisotropic medium. We carry out rigorous 3D simulations for both the conventional and the sub-wavelength MMI in section 3, to assess their optimum performance. The experimental results are presented in section 4, 

\section*{Results}
\textbf{Self-imaging revisted}.
We first focus on the 2D model of the MMI shown in Fig. \ref{fig:mmi_conv} and \ref{fig:mmi3d} in order to compare self-imaging in conventional and sub-wavelength engineered MMIs. In this 2D model the $y$-axis is contracted using the effective index method \cite{ChenGuided}. In both devices the mode field launched from the input waveguide,  $f_\mathrm{in}(x)$, is expanded into the modes of the multimode slab region, i.e. $f_\mathrm{in}(x)=\sum_{i} c_i \varphi_i(x)$, with $c_i$  the overlap between the input field and the $i$-th slab mode.
These modes propagate with  specific phase constants, $\beta_i$, producing self-images of the input field as they interfere. The positions at which these images form is governed by the beat length of the two lowest order modes \cite{soldano1995optical}, 
\begin{equation}
L_\pi = \frac{\pi}{\beta_1-\beta_2}. 
\label{eq:Lpi}
\end{equation} 
For instance, a double image of the input field is  formed at $z=\frac{3}{2}L_\pi=L_\mathrm{MMI}$, where $L_\mathrm{MMI}$ is the physical length of the multimode region (see Fig. \ref{fig:mmi_conv}). Since $L_\mathrm{MMI}$ is fixed, any wavelength variation of the beat length will result in a detuning of the device and in a significant degradation of its performance.  

\textbf{Conventional multimode interference.}
In a conventional MMI, under the paraxial approximation,  the beat length is  given by 
\begin{equation}
L_\pi^\mathrm{conv} \approx \frac{4  W_\mathrm{e}^2}{3\lambda}n_\mathrm{core},
\label{eq:LpiSoldano}
\end{equation} 
where $n_\mathrm{core}$ is the effective index of the multimode region obtained with the effective index method \cite{ChenGuided} and $\lambda$ is the free space wavelength.  $W_e$  is the effective width of the multimode region taking into account the Goos-H\"anchen shift, which is assumed to be identical for all modes and invariant with wavelength  \cite{soldano1995optical}. 
The choice of the material platform clearly dictates the value of $n_\mathrm{core}$ and its wavelength dependence. The minimum width of the multimode region is determined by the separation ($s$) between the input/output waveguides, to avoid coupling between them, and the required width of the access waveguides ($W_A$), to control the excitation of higher order modes; see \cite{halir2008design,Halir:11} for details. Hence, there is very limited freedom to engineer the beat length or its wavelength dependence.  
\begin{table}[tbp]
\centering
\caption{Geometrical parameters of the conventional and the broadband MMI shown in Fig. \ref{fig:mmi_conv} and \ref{fig:mmi3d}, respectively.}
\begin{tabular}{|c|c|c|}
\hline
Parameter & Conventional & Broadband \\
\hline \hline
Silicon thickness ($H$) & \multicolumn{2}{c|}{$220\,\mathrm{nm}$}  \\ \hline
Width MMI ($W_\mathrm{MMI}$)& \multicolumn{2}{c|}{$3.25\,\mu\mathrm{m}$}  \\ \hline
Length MMI ($L_\mathrm{MMI}$) & $38.5\,\mu\mathrm{m}$ & $
\begin{array}{c}
14\,\mu\mathrm{m} \\
74\,\mathrm{periods}\\
\end{array}$  \\ \hline
Width input ($W_I$) & \multicolumn{2}{c|}{$0.5\,\mu\mathrm{m}$}  \\ \hline
Width access ($W_A$) & \multicolumn{2}{c|}{$1.7\,\mu\mathrm{m}$} \\ \hline
Length taper ($L_T$) & $6\,\mu\mathrm{m}$ & $
\begin{array}{c}
5.7\,\mu\mathrm{m} \\
30\,\mathrm{periods}\\
\end{array}$  \\ \hline
Separation ($s$) & \multicolumn{2}{c|}{$0.3\,\mu\mathrm{m}$} \\ \hline
Period ($\Lambda$) & -- & $190\,\mathrm{nm}$ \\ \hline
Duty cycle (DC) & -- & 50\,\% \\ \hline
$n_{\mathrm{core}}$ @ $\lambda=1500\,\mathrm{nm}$  & $\sim 2.87$ & 
$\begin{array}{c}
n_{xx} \sim 2.15 \\
n_{zz} \sim 1.6\\
\end{array}$\\
\hline
\end{tabular}
  \label{tab:dimensions-mmi}
\end{table}

\begin{figure}[t]
\centering
{\includegraphics[width=\linewidth]{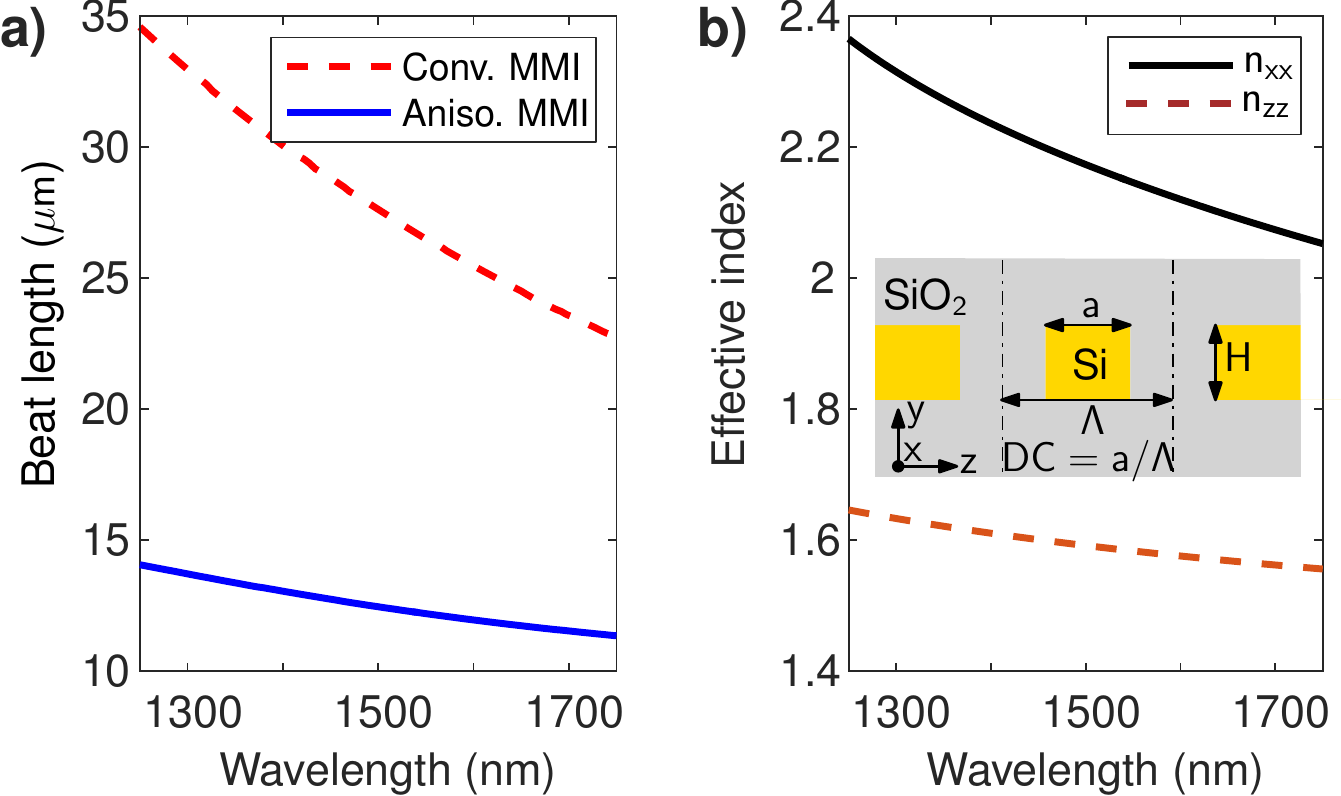}}
\caption{a) Semi-analytic calculation of the beat length of a conventional MMI device and the sub-wavelength engineered MMI modeled through the anisotropic medium $\mathbf{n_{core}}=\mathrm{diag}[n_{xx},n_{zz}]$.  The anisotropy of the sub-wavelength engineered MMI yields an almost threefold reduction in beat length compared to the conventional MMI, as well as a substantially reduced wavelength dependence. b) Anisotropic effective indexes of the sub-wavelength grating metamaterial waveguide as a function of wavelength as obtained through modal analysis from the structure shown in the inset,  for a duty-cycle of $\mathrm{DC}=50\%$ and a period of $\Lambda=190\,\mathrm{nm}$.} 
\label{fig:nir_analytical_lpi}
\end{figure}

Considering a $H=220\,\mathrm{nm}$ thick silicon platform with silicon dioxide ($\mathrm{SiO_2}$) upper cladding, TE polarization, and the geometrical parameters given in table \ref{tab:dimensions-mmi}, Eq. (\ref{eq:LpiSoldano}) yields   $L_\pi^\mathrm{conv} \approx 27\,\mu\mathrm{m}$ at $\lambda=1500\,\mathrm{nm}$. Figure \ref{fig:nir_analytical_lpi}(a) shows the  beat length  calculated through Eq. (\ref{eq:Lpi}), with $\beta_{1,2}$ obtained from a slab waveguide of width $W_\mathrm{MMI}$,  silicon dioxide cladding, and effective core index $n_\mathrm{core}$.  The beat length is $L_\pi^\mathrm{conv}\sim 28\,\mu\mathrm{m}$ at $\lambda=1500\,\mathrm{nm}$,  in good agreement with Eq. (\ref{eq:LpiSoldano}). However, due to the  strong wavelength dependence, the device will only operate in a limited bandwidth of $\sim 200\,\mathrm{nm}$ as discussed below.

\textbf{Multimode interference in anisotropic media.}
The fundamental advantage of our broadband MMI device, shown in Fig. \ref{fig:mmi3d}, arises from the inherent anisotropy of its sub-wavelength structure: two guided waves traveling in the central multimode region along the $z$ and $x$ directions experience optically different structures \cite{bouchitte1985homogenization,gu1996form}. Referring to the effective indexes of these waves as $n_{xx}$ and $n_{zz}$, respectively, we see that in the 2D model shown in Fig. \ref{fig:mmi_conv} the equivalent material in the multimode region can be described by the diagonal tensor $\mathbf{n_{core}}=\mathrm{diag}[n_{xx},n_{zz}]$. In such an anisotropic medium, the dispersion equation is given by $(k_x/n_{zz})^2+(k_z/n_{xx})^2=k_0^2$, where $k_x$ is the wave-vector component in the $x$ direction, $k_z=\beta$ is the propagation constant,  and $k_0=2\pi/\lambda$ \cite{satomura1974}. Extending the procedure outlined in \cite{soldano1995optical} for conventional isotropic MMIs, we find that in the anisotropic case the beat length is given by:
\begin{equation}
L_\pi^\mathrm{aniso} \approx \frac{4  W_\mathrm{e}^2}{3\lambda} \frac{n_{zz}^2}{n_{xx}}.
\label{eq:LpiHalir}
\end{equation} 
Equation (\ref{eq:LpiHalir}) provides an analytical design framework for our ultra-broadband MMIs.

The elements of effective index tensor can be calculated from the structure shown in the inset of Fig. \ref{fig:nir_analytical_lpi}(b): $n_{xx}$ is the effective index of the fundamental Bloch mode traveling in the $z$ direction and polarized along the $x$ axis; $n_{zz}$ is the effective index of the fundamental Bloch mode traveling in the $x$ direction and polarized along the $z$ axis. For the dimensions given in table \ref{tab:dimensions-mmi}, we obtain the indexes shown in Fig. \ref{fig:nir_analytical_lpi}(b); specifically  at $\lambda=1500\,\mathrm{nm}$ we have $ n_{xx} \sim 2.15$, $ n_{zz} \sim 1.6$. Equation (\ref{eq:LpiHalir}) then predicts a significantly reduced beat length of $L_\pi^\mathrm{aniso} \approx 11.5\,\mu\mathrm{m}$,  which arises directly from the reduced value of $n_{zz}^2/n_{xx}\sim 1.2$ compared to $n_\mathrm{core}\sim 2.87$ [see Eqs. (\ref{eq:LpiSoldano}) and (\ref{eq:LpiHalir})].  Since the variation of $L_\pi^\mathrm{aniso}$ and $L_\pi^\mathrm{conv}$
with wavelength is, to first order, proportional to $n_{zz}^2/n_{xx}$ and  $n_\mathrm{core}$, the wavelength dependence of the beat length is  expected to decrease, too.
Figure \ref{fig:nir_analytical_lpi}(a) shows the beat length calculated through Eq. (\ref{eq:Lpi}), with $\beta_{1,2}$ obtained from an anisotropic slab waveguide of width $W_\mathrm{MMI}$, silicon dioxide cladding, and $\mathbf{n_{core}}=\mathrm{diag}[n_{xx},n_{zz}]$  \cite{satomura1974}.  From the figure, we find $L_\pi^\mathrm{aniso} \sim 12\,\mu\mathrm{m}$ at $\lambda=1500\,\mathrm{nm}$, in good agreement with  Eq. (\ref{eq:LpiHalir}),  and observe that the beat length becomes significantly flatter with wavelength. We furthermore found that the beat length is minimum for a duty cycle of $\sim 50\,\%$, which facilitates fabrication, as it results in the largest linewidths for a given grating period. 
It is noted that the 
precise dispersion behavior of the MMI modes near their Bragg wavelength can only be obtained in full 3D simulations, which enables us to further flatten the beat length by fine-tuning the pitch of the structure.

\begin{figure}[tbp]
\centering
{\includegraphics[width=\linewidth]{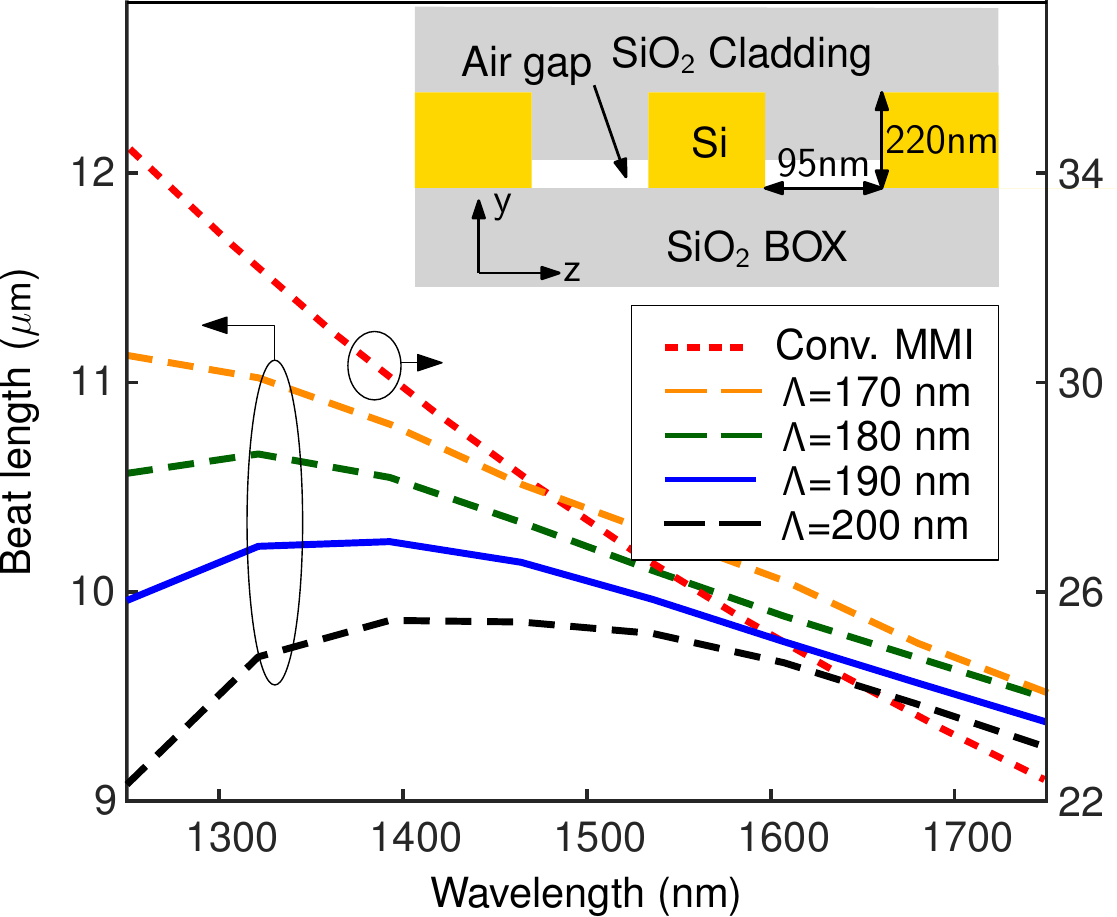}}
\caption{Full vectorial 3D simulations of beat length  as a function of wavelength for a conventional MMI device (right scale) and the broadband MMI (left scale).   With a judiciously designed pitch, the modal dispersion of the SWG structure in the broadband MMI further reduces the wavelength dependence of the beat length. Inset: Schematic side-view of the SWG in the broadband MMI showing small air gaps under the $\mathrm{SiO}_2$ upper cladding. The geometric parameters of the device are given in table \ref{tab:dimensions-mmi}.}
\label{fig:nir_lpi}
\end{figure}

\begin{figure}[t]
\centering
{\includegraphics[width=\linewidth]{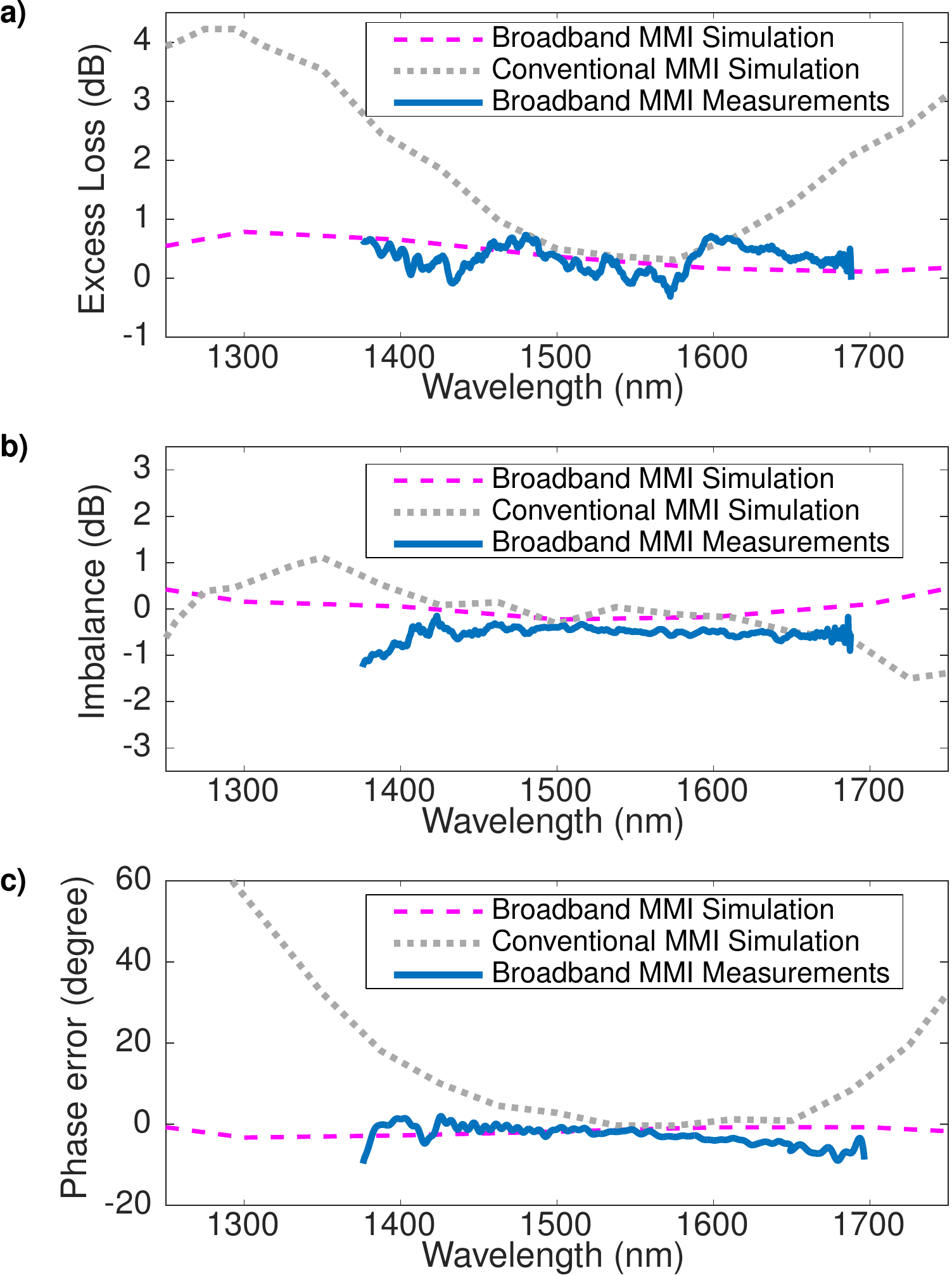}}
\caption{Measured and simulated performance of the broadband MMI compared to  an optimized conventional MMI design, including a) excess loss, b) imbalance and c) phase error. The broadband coupler design shows virtually ideal performance over a  $500\,\mathrm{nm}$  bandwidth and the fabricated device has a measured bandwidth of  $300\,\mathrm{nm}$, while the conventional MMI design has a bandwidth of less  than  $200\,\mathrm{nm}$ and the footprint is  three times larger.}
\label{fig:nir_performance}
\end{figure}

\textbf{Design and simulation.}
We use full 3D simulations to assess and further optimize the performance of both the conventional and the sub-wavelength metamaterial engineered MMI.  Figure \ref{fig:nir_lpi} shows the simulated wavelength dependence of the beat length of a conventional MMI; the propagation constants $\beta_{1,2}$ were obtained with a commercial mode solver \cite{FIMM}. As expected from the 2D model [Fig. \ref{fig:nir_analytical_lpi}(a)], the beat length exhibits a strong variation with wavelength, between $35\,\mu\mathrm{m}$ and $23\,\mu\mathrm{m}$, resulting in performance degratation when  de-tuning the device from its design wavelenght.  For a conventional MMI with the optimized dimensions shown in table \ref{tab:dimensions-mmi}, the simulated impact of this de-tuning is shown Fig. \ref{fig:nir_performance}, including the excess losses (EL), imbalance (IB) and phase error (PE). By denoting  the complex  transmission from the fundamental mode of input waveguide 1 to the fundamental mode of the output waveguides 3 and 4 as $s_{31}$ and $s_{41}$  (see Fig. \ref{fig:mmi_conv}), these performance parameters are: $\mathrm{EL}=10\log\left(|s_{31}|^2+|s_{41}|^2\right)$, $\mathrm{IB}=10\log\left(|s_{31}|^2/|s_{41}|^2\right)$, and $\mathrm{PE}=\angle (s_{31}/s_{41})-90^\circ$. From Figs. \ref{fig:nir_performance}(a)-(c) it is obvious that as the device is operated further away from its design wavelength its performance steadily deteriorates. Aiming for excess losses and imbalance below $1\,\mathrm{dB}$ and phase error smaller than $5^\circ$ thus yields a bandwidth of  under $200\,\mathrm{nm}$ for this optimized  MMI design. 

The wavelength dependence of the beat length in our sub-wavelength engineered MMI (see Fig. \ref{fig:mmi3d}) is shown, for a $50\,\%$ duty cycle, in Fig. \ref{fig:nir_lpi}; the propagation constants $\beta_{1,2}$ were obtained using 3D Finite Difference Time Domain (FDTD) simulations \cite{RSOFT}, and the procedure described in \cite{GonzaloWanguemert-Perez:14}. For the device parameters given in table \ref{tab:dimensions-mmi},  we find that by fine-tuning the pitch near $190\,\mathrm{nm}$, and thereby adjusting the  dispersion of the MMI modes, the beat length becomes virtually wavelength independent in the $1250\,\mathrm{nm}$ to $1750\,\mathrm{nm}$ wavelength range. Furthermore, the beat length at $\lambda=1500\,\mathrm{nm}$ is $L_\pi^\mathrm{SWG} \sim 10\,\mu\mathrm{m}$, which is a threefold reduction compared to the conventional device.  The inset of Fig. \ref{fig:nir_lpi} illustrates the longitudinal cross-section of the MMI. In our initial design we assumed that the upper $\mathrm{SiO_2}$ would completely fill the trenches between the silicon segments. However, our experimental results suggest that this is not the case, and that in fact an air gap of approximately $60\,\mathrm{nm}$ is left, due to the  relatively high  aspect ratio of the trenches. All 3D simulation results presented here include this gap.

For a pitch of $\Lambda = 190\,\mathrm{nm}$ the wavelength average beat length is $L_\pi \approx 10\,\mu\mathrm{m}$, yielding a device with approximately $(3/2) L_\pi / \Lambda = 79\,\mathrm{periods}$ for a $2\times 2$  MMI device. We optimized the length of the device using 3D-FDTD simulations of the complete structure that take into account material dispersion. Device performance for the optimized length of $74\,\mathrm{periods}$ is shown in Fig. \ref{fig:nir_performance}. These results confirm virtually perfect device operation over a $500\,\mathrm{nm}$ wavelength span, with excess losses and imbalance below $1\,\mathrm{dB}$, and phase error less than $5^\circ$. This is an unprecedented almost threefold bandwidth enhancement compared to the conventional MMI.   

\begin{figure}[t]
\centering
{\includegraphics[width=\linewidth]{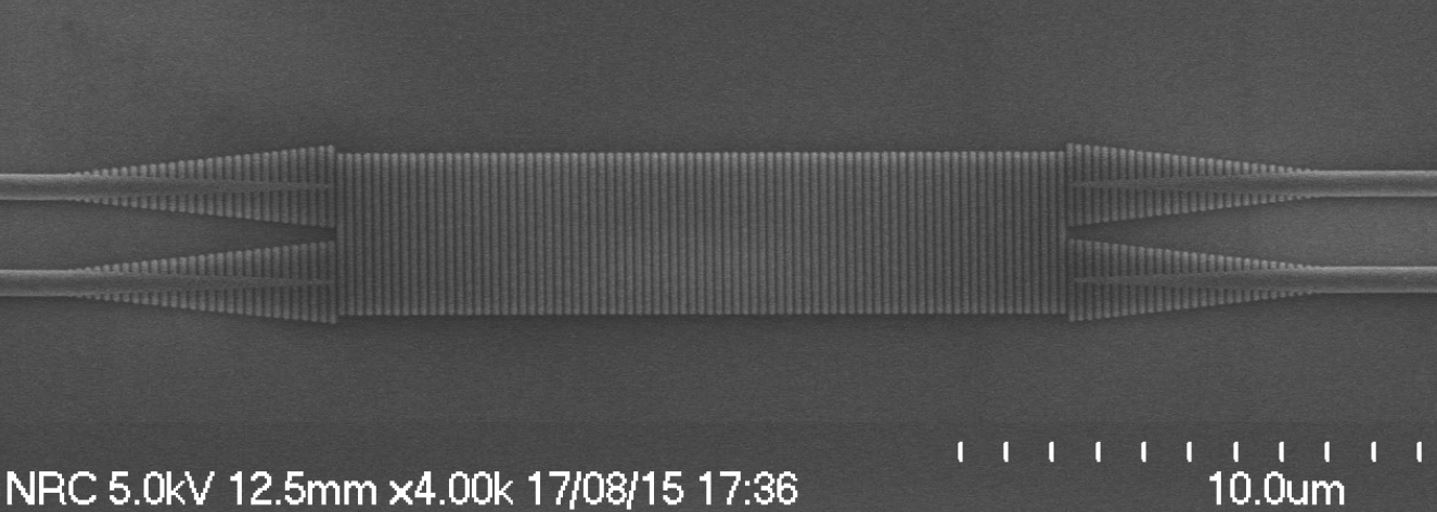}}
\caption{Scanning electron microscope image of a  fabricated ultra-broadband MMI, prior to the deposition of the $\mathrm{SiO_2}$ cladding.}
\label{fig:nir_sem}
\end{figure}

\textbf{Experimental results.}
In order to experimentally validate the broadband operation of our device, several test structures were fabricated on a standard $220\,\mathrm{nm}$ silicon-on-insulator wafer, as described in the methods.  Figure  \ref{fig:nir_sem} shows a Scanning Electron Microscope (SEM) image of one of the fabricated devices prior to the deposition of the $\mathrm{SiO_2}$ upper cladding. Measurements of the devices were carried out on two independent setups (see methods). The experimental results shown in Fig. \ref{fig:nir_performance} reveal  virtually perfect device performance in the $1375\,\mathrm{nm}$ -  $1700\,\mathrm{nm}$ wavelength range: excess losses and imbalance are below $1\,\mathrm{dB}$, and the phase error is smaller than $5^\circ$, thereby confirming the broadband behavior of the device. We observed a resonant behavior near $1350\,\mathrm{nm}$, which is attributed to the increased duty cycle of the fabricated devices and could be addressed with proper pre-scaling of the mask. Our broadband source does not extend beyond $1700\,\mathrm{nm}$ preventing measurements at longer wavelengths.

\section*{Discussion}
In conclusion, we have shown that the intrinsic anisotropy of sub-wavelength engineered metamaterials can be exploited to design compact and ultra-broadband nanophotonic beam splitters.  Specifically, by controlling anisotropy and dispersion of a nanophotonic metamaterial waveguide, we have demonstrated MMI designs with virtually ideal performance over a $500\,\mathrm{nm}$ and have fabricated a device with a measured bandwidth in excess of $300\,\mathrm{nm}$, outperforming conventional MMIs by a wide margin.
 We believe that this new strategy to engineer ultra-broadband waveguide devices can usher in broadband integrated nanophotonic systems with applications in coherent communications, sensing, spectroscopy and quantum photonics.   

\section*{Methods}
\textbf{Fabrication.}  We fabricated a chip containing both individual MMIs, asymmetric Mach-Zehnder interferometers and reference waveguides. To enable broadband light coupling to the chip we used sub-wavelength engineered fiber-to-chip mode transformers \cite{Cheben:15}.  The devices were defined with electron beam lithography and transferred to the silicon layer with reactive ion etching. The  $\mathrm{SiO_2}$ upper cladding was deposited using plasma-enhanced chemical vapor deposition. 

\textbf{Measurements.} The devices were characterized with two independent setups. An initial screening of the devices was carried out using a tunable laser source at the input and power-detector at the output, covering the $1400\,\mathrm{nm}$ to  $1600\,\mathrm{nm}$ range. In order to measure in an even broader range,  we used a second setup with a broadband light source as input and an optical spectrum analyzer at the output. In both cases the input polarization was set to quasi-TE (electric field in the horizontal plane) using a polarization controller. To determine excess losses, the transmission through an individual MMI was normalized to the  transmission through a reference waveguide. The phase error was obtained from the interferogram produced by the asymmetric Mach-Zehnder interferometers, using a minimum-phase technique \cite{halir2009characterization}.  The measurement data from both setups was found to be in very good agreement in the common wavelength range, with a deviation below $5^\circ$ in phase, $0.5\,\mathrm{dB}$ in imbalance and $1\,\mathrm{dB}$ in excess loss.

\bibliographystyle{naturemag}
\bibliography{mmi_bib}

\section*{Acknowledgements}
We acknowledge funding from the Ministerio de Economía y Competitividad, Programa Estatal de Investigación, Desarrollo e Innovación Orientada a los Retos de la Sociedad (cofinanciado FEDER), Proyecto TEC2013-46917-C2-1-R. We would like to thank Dr. Jean Lapointe for preparing files for e-beam lithography.

%\section*{Author contributions}
%R.H. conceived the original concept, pre-designed the devices, carried out part of the measurements, wrote the manuscript and supervised additional simulation work. P.C. contributed to writing the manuscript, and helped design the test structures. J.M.L.G. and J.D.S.M. carried out  simulations related to device optimization and anisotropic modeling. J.H.S and D.X.X. provided assessment on the fabrication of the structure and contributed to writing the manuscript. G.W.P., I.M.F. and A.O.M. worked with R.H. on the original concept, provided guidelines for the pre-design of the devices, assessment on the anisotropic modeling and helped prepare the manuscript. S.W. carried out part of the measurements. All authors contributed to interpreting the data and editing the manuscript.

%\section*{Competing financial interests}
%The authors declare no competing financial interests.

\end{document}